\documentclass{aa}  
\usepackage{orcidlink}
\usepackage{comment}
\usepackage{graphicx}
\usepackage{txfonts}

\usepackage{natbib}
\bibpunct{(}{)}{;}{a}{}{,}

\begin{document} 

\title{The position of SN~1987A}

\subtitle{}

\author{Christos Tegkelidis\inst{1}\orcidlink{0009-0007-3359-5767} \and Josefin Larsson\inst{1}\orcidlink{0000-0003-0065-2933} \and Dennis Alp\inst{1}\orcidlink{0000-0002-0427-5592}
        }

\institute{Department of Physics, KTH Royal Institute of Technology, The Oskar                Klein Centre, AlbaNova, SE-106 91 Stockholm, Sweden\\
\email{cteg@kth.se}
             }

\date{Received }

  \abstract 
   {}
   {The accurate positional measurement of Supernova (SN) 1987A is important for determining the kick velocity of its compact object and the velocities of the ejecta and various shock components. In this work, we perform absolute astrometry to determine the position of SN~1987A.}
   {We used multi--epoch Hubble Space Telescope imaging to model the early ejecta and the equatorial ring (ER). We combined our measurements and obtained the celestial coordinates in the International Celestial Reference System (ICRS) by registering the observations onto Gaia Data Release 3.}
   {The final average position of the different measurements is ${\alpha = 5^{\mathrm{h}}~ 35^{\rm{m}}~ 27^{\rm{s}}.9884(30)}$, ${\delta = -69^{\circ}~ 16'~ 11''.1134(136)}$ (ICRS J2016). The early ejecta position is located 14~mas south and 16~mas east of the ER center, with the offset being significant at 96\% confidence. The offset may be due to instrument and/or filter--dependent systematics and registration uncertainties, though an intrinsic explosion offset relative to the ER remains possible. Image registration with proper motion corrections yields similar astrometry and a source proper motion of ${\mu_{\rm east} (\equiv \rm{PM_{\alpha }*}) = 1.60 \pm 0.15 ~\rm{mas ~ yr^{-1}}}$ and ${\mu_{\rm{north}} (\equiv \rm{PM_{\delta}}) = 0.44 \pm 0.09~\rm{mas ~ yr^{-1}}}$, in agreement with the typical local motion of the Large Magellanic Cloud.}
   {The absolute positional uncertainty of 21~mas adds a systematic uncertainty to the sky--plane kick velocity of ${123}~(t/40~\rm{yr})^{-1}~\rm{km~s}^{-1}$, where $t$ is the time since the explosion. Comparing the location of the compact source observed with JWST to our updated position implies a sky--plane kick of ${399\pm148~\mathrm{km~s^{-1}}}$ and a 3D kick of ${472\pm126~\mathrm{km~s^{-1}}}$, which is consistent with previous estimates.}
    
   
   \maketitle

\section{Introduction}

Supernova (SN) 1987A is located in the Large Magellanic Cloud (LMC) at a distance of 49.6~kpc \citep{2019Natur.567..200P}. It has been regularly observed across the electromagnetic spectrum, which has led to a detailed understanding of the properties of the explosion and subsequent evolution into the remnant phase (see \citealp{2016ARA&A..54...19M} for a review). At current epochs, the emission from the system is dominated by the shock interaction between the ejecta and the circumstellar equatorial ring (ER), which was ejected from the progenitor in a mass--loss episode $\sim 20,000$~years before the explosion \citep{2000ApJ...528..426C}.

Despite extensive studies of the remnant, there is an uncertainty in its precise position. Determining the absolute position of SN~1987A is essential to connect observations across multiple wavebands and constrain the projected sky velocities of the freely expanding ejecta and the various shock components. Moreover, in core-collapse SNe, newly formed neutron stars are thought to acquire kick velocities as a result of explosion asymmetries, which may include uneven mass ejection and anisotropic neutrino emission \citep[e.g.,][]{2024Ap&SS.369...80J}. Therefore, accurate astrometry can benefit future studies quantifying the kick of SN~1987A's neutron star, which is especially relevant in view of the recent discovery of emission associated with the compact object with the James Webb Space Telescope (JWST) \citep{2024Sci...383..898F}.

Early attempts to determine its precise location were based on radio observations with the Australia Telescope Compact Array and optical observations with the Hubble Space Telescope (HST)  \citep{1995A&A...304..116R}. However, these initial studies were limited by the lack of accurate image registration. More recent investigations have focused on the analysis of the ER in the optical \citep{2018ApJ...864..174A} and radio domain \citep{2019ApJ...886...51C} using HST and Atacama Large Millimeter Array (ALMA) observations, respectively. In this work, we combine a large set of HST observations in various bands and from different epochs to determine the position of SN~1987A. We address systematic uncertainties by analyzing the remnant in different evolutionary stages. We register our results onto the latest Gaia Data Release 3 (DR3) \citep{2016A&A...595A...1G, 2023A&A...674A...1G}, which provides highly precise astrometric and photometric measurements for nearly two billion celestial objects. 

This paper is organized as follows. We present the observations in Sect.~\ref{Observations} and then explain the methods for position estimation and image registration in Sect.~\ref{Methods}. The results are presented in Sect.~\ref{Results}, followed by a discussion in Sect.~\ref{Discussion}.

\section{Observations}\label{Observations}

\begin{table*}
\caption{HST /FOC observations}
\label{Tab: Observations} 
\centering 
\begin{tabular}{c c c c c c c} 
\hline\hline 
Obs. ID &  Date & Epoch\tablefootmark{a} & Filter & Exposure & Center wavelength & Stars\tablefootmark{b} \\ 
 & YYYYmmdd & (days) & & (s) & (Å)\\
\hline 
X0C80102T & 1990-08-24 & 1278 & F175W & 838 & 1720 & 2, 3, 6\\
X0C80103T & 1990-08-24 & 1278 & F175W & 838 & 1720 & 2, 3, 6\\
X0C80106T & 1990-08-24 & 1278 & F501N & 822 & 5010 & 2, 3, 4, 5, 6 \\
X0C80107T & 1990-08-24 & 1278 & F501N & 838 & 5010 & 2, 3, 4, 6 \\
\hline 
\end{tabular}
\tablefoot{\\
\tablefoottext{a}{Since 1987-02-23.}
\tablefoottext{b}{Stars used for image registration, see Fig.~\ref{Fig: Observations}.}
}
\end{table*}

\begin{table*}
\caption{HST /WFPC2, ACS/HRC, and WFC3/UVIS observations} 
\label{Tab: All Observations} 
\centering 
\begin{tabular}{c c c c c c c c c c}
\hline\hline 
Date & Epoch\tablefootmark{a} & Instrument & Filter & Exposure &Date & Epoch\tablefootmark{a} & Instrument & Filter & Exposure\\
YYYYmmdd& (days) & & & (s) & YYYYmmdd & (days) & & & (s) \\
\hline
1994-09-24 & 2770 & WFPC2   & F675W & 600  & 2007-05-12 & 7384  & WFPC2      & F675W & 2700 \\
1995-03-05 & 2932 & WFPC2   & F675W & 600  & 2008-02-19 & 7666  & WFPC2      & F675W & 1600 \\
1996-02-06 & 3270 & WFPC2   & F675W & 600  & 2009-04-29 & 8101  & WFPC2      & F675W & 1600 \\
1997-07-10 & 3790 & WFPC2   & F675W & 600  & 2009-12-12 & 8328  & WFC3/UVIS  & F625W & 3000 \\
1998-02-06 & 4001 & WFPC2   & F675W & 400  & 2011-01-05 & 8717  & WFC3/UVIS  & F625W & 1140 \\
1999-01-07 & 4336 & WFPC2   & F675W & 1220 & 2013-02-06 & 9480  & WFC3/UVIS  & F625W & 1200 \\
1999-04-21 & 4440 & WFPC2   & F675W & 400  & 2014-06-15 & 9974  & WFC3/UVIS  & F625W & 1200 \\
2000-02-02 & 4727 & WFPC2   & F675W & 400  & 2015-05-24 & 10317 & WFC3/UVIS  & F625W & 1200 \\
2000-06-16 & 4862 & WFPC2   & F675W & 400  & 2016-06-08 & 10698 & WFC3/UVIS  & F625W & 600 \\
2000-11-13 & 5013 & WFPC2   & F675W & 2400 & 2017-08-03 & 11119 & WFC3/UVIS  & F625W & 1200 \\
2001-03-23 & 5142 & WFPC2   & F675W & 500  & 2018-07-08 & 11458 & WFC3/UVIS  & F625W & 1200 \\
2001-12-07 & 5401 & WFPC2   & F675W & 800  & 2019-07-22 & 11837 & WFC3/UVIS  & F625W & 1200 \\
2003-01-05 & 5796 & ACS/HRC & F625W & 800  & 2020-08-06 & 12218 & WFC3/UVIS  & F625W & 1160 \\
2003-08-12 & 6014 & ACS/HRC & F625W & 480  & 2021-08-21 & 12598 & WFC3/UVIS  & F625W & 1080 \\
2003-11-28 & 6122 & ACS/HRC & F625W & 800  & 2022-09-05 & 12978 & WFC3/UVIS  & F625W & 1080 \\
2005-09-26 & 6790 & ACS/HRC & F625W & 7200 \\
2006-04-15 & 6991& ACS/HRC  & F625W & 1200 \\
2006-12-06 & 7226 & ACS/HRC & F625W & 1200 \\
\hline
\end{tabular}
\tablefoot{\\
\tablefoottext{a}{Since 1987-02-23.}
}
\end{table*}

\begin{table*}
\caption{Stars used for the astrometric registration of the FOC images}
\label{Tab: star registration}
\centering
\begin{tabular}{c c c c c c c}
    \hline\hline 
    Star & Source ID & RA  & $\sigma_{RA}$ & Dec & $\sigma_{Dec}$ & G \\
    & & (°)&(mas)&(°)&(mas)& (mag)\\
    \hline 
    2 & 4657668080079343104 & 83.86512285 & 0.0241 & -69.26915037 & 0.0244 & 14.991971 \\
    3 & 4657668007030021248 & 83.86777270 & 0.0324 & -69.26996955 & 0.0326 & 15.773073 \\
    4 & 4657668080057143424 & 83.86514334 & 0.2260 & -69.27002360 & 0.4634 & 19.306423 \\
    5 & 4657668007091887360 & 83.86955760 & 0.8538 & -69.26923336 & 0.7894 & 20.61967 \\ 
    6 & 4657668075811272704 & 83.86242898 & 0.1085 & -69.27017562 & 0.1122 & 18.204403 \\
    \hline
\end{tabular}

\tablefoot{Information obtained from Gaia DR3. The stars are labeled in accordance to Fig.~\ref{Fig: Observations} and \citet{1987ApJ...321L..41W}.\\
}
\end{table*}

\begin{figure*}
\centering
\includegraphics[width=17cm]{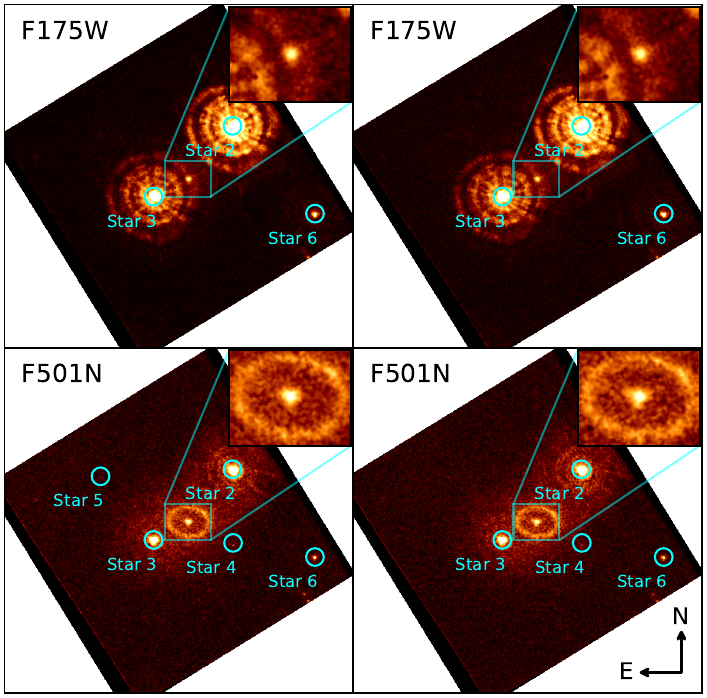}
\caption{HST/FOC imaging observations of SN~1987A. The color scale of the images is inverse hyperbolic sine and arbitrarily scaled to highlight the ejecta in the center. For visualization only, the images are resampled to a common grid where north is up and east is left. The field of view (FOV) of each full image is $11'' \times 11''$; zoomed-in insets focus on the expanding ejecta and cover $1\farcs950  \times 1\farcs525$.}
\label{Fig: Observations}
\end{figure*}

The analysis is based on four early observations obtained with the HST/Faint Object Camera (FOC) and 33 later observations obtained with HST's Wide Field and Planetary Camera 2 (WFPC2), Advanced Camera for Surveys (ACS) and Wide Field Camera 3 (WFC3) instruments. We provide an overview of the early FOC observations in Table \ref{Tab: Observations} and the later observations in Table \ref{Tab: All Observations}. The observations in Table \ref{Tab: All Observations} have also been discussed in \cite{2011Natur.474..484L, 2015ApJ...806L..19F, 2019ApJ...886..147L, 2024ApJ...966..238R}.

The FOC images consist of two exposures in the broad--band F175W UV filter and two exposures in the narrow--band F501N [O III] $\lambda5007$ filter (see Fig.~\ref{Fig: Observations} and Table~\ref{Tab: Observations}). These observations were obtained in the year 1990, only 1278 days after the explosion. We excluded two exposures from the same epoch in the broad-band filters F275W and F346M, as these images exhibit significant geometric distortions due to saturation in the two brightest stars (identified as stars 2 and 3 in \citealt{1987ApJ...321L..41W}). We downloaded the pipeline--processed images from the Mikulski Archive for Space Telescopes (MAST). All observations were conducted with the f/96 camera mode and have a pixel scale of 22~mas~pixel$^{-1}$ and a field--of--view (FOV) of ${11'' \times 11''}$. These observations were conducted before the installation of the Corrective Optics Space Telescope Axial Replacement (COSTAR), which corrected the spherical aberration of the primary mirror. This aberration causes broad wings in the PSF, significantly blurring extended sources. However, the position of the PSF center should remain unaffected.

The WFPC2 images were obtained with the Planetary Camera (PC1) in the F675W filter, the ACS observations used the High--Resolution Channel (HRC) with the F625W filter, while the WFC3 observations used the UVIS detector with the F625W filter (see Table \ref{Tab: All Observations}). For simplicity, we refer to all these filters as the \textit{R}--band. The FOV of the images is $34'' \times 34''$ for WFPC2/PC1, $26'' \times 29''$ for ACS/HRC, $20\farcs5 \times 20\farcs5$ for WFC3/UVIS on day 8328, 8717, 10,698, and 11,119, and $41''\times41''$ for all other WFC3/UVIS observations. All observations after 5401 days were obtained with a four--point dither pattern, except the WFPC2 observation on day 7384, which was undithered. All other observations were undithered.

All individual exposures were combined with \texttt{DrizzlePac} \citep{2021AAS...23821602H}, using the Hubble Advanced Products Single--Visit Mosaics retrieved from the MAST, which provide the best relative alignment within visits. Where supported by the number and geometry of the dithers, we over--sampled the images to improve the resolution by optimizing the final pixel scale and the \texttt{final\_pixfrac} parameters.
The optimization was based on visual inspection of the images and analysis of the drizzle weight maps and background noise in the weight images, following the recommendations of the \texttt{DrizzlePac} handbook \citep{DrizzlePacHandbook2025}. For WFPC2 products, we retained the native plate scale of \(46~\mathrm{mas~pixel^{-1}}\) as most observation groups contained only a few exposures with limited dithering. The only exceptions were three observations, which both employed a four--point dithering pattern; one on day 5401 with a final pixel scale of 25~mas and a \texttt{final\_pixfrac} of 0.8, one on day 7666 with a final pixel scale of 25~mas and a \texttt{final\_pixfrac} of 0.8, and one on day 8101 with a final pixel scale of 25~mas and a \texttt{final\_pixfrac} of 0.6 (see Table \ref{Tab: All Observations}). For ACS/HRC, we retained the native plate scale of \(25~\mathrm{mas~pixel^{-1}}\), since these observations are critically sampled at 630~nm \citep{2022acsd.book...11L}. For WFC3/UVIS, we adopted a final pixel scale of 25~mas with a \texttt{final\_pixfrac} of 0.7. All of the above configurations provided the best resolution without a significant increase in the noise and a ratio of background root mean square (RMS) to weight--predicted noise less than 0.15 in all cases.

\section{Methods}\label{Methods}

\subsection{Celestial reference system}\label{Subsection: Celestial Reference System}

All reported coordinates are based on the International Celestial Reference System (ICRS), the standard celestial reference system established by the International Astronomical Union (IAU) as of January 1, 1998. The ICRS is centered at the barycenter of the solar system, with axes that remain fixed relative to distant extragalactic objects. Its current realization at radio wavelengths is the International Celestial Reference Frame (ICRF), which consists of precisely measured positions of extragalactic radio sources that define the orientation of the system. Although the ICRS axes are closely aligned with the mean equator and equinox of the J2000 frame, they are not tied to Earth's motion. As a result, the orientation of the system is independent of epoch, and thus coordinates reported in frames that are realizations of the ICRS do not require an associated equinox \citep{2005USNOC.179.....K}. However, an epoch must still be specified when reporting coordinates for objects with non-negligible proper motion, as their apparent positions change over time. Throughout this work, we adopted the Gaia~DR3 reference epoch of J2016 when using Gaia--based astrometry.

\subsection{Position estimation}\label{Subsection: Position Estimation}

We determined the position of SN~1987A assuming that it coincides with the center of the early ejecta and the center of the ER. The two methods provide independent measurements of the center of the explosion, enabling us to assess the systematic uncertainties associated with the underlying assumptions. In the case of the ejecta, we assume that the peak brightness at early times traces the center of the explosion. For the ER, we instead assume that the progenitor, and hence also the explosion, is located at its center. 

\subsubsection{Early ejecta}\label{Subsection: Early Ejecta}

We analyzed the early emission of the marginally resolved ejecta in the FOC images. The bulk of the ejecta that dominate the emission have not expanded significantly from the position of the progenitor star at these times. In addition, we find no sign of asymmetries in the marginally resolved images. We modeled the early ejecta using two co--spatial 2D elliptical Gaussians with the same orientation but different amplitudes and widths to capture both the narrow core and broad wings of the PSF. We used the covariance matrix of the fits as an estimate for the position uncertainties of the ejecta.

\subsubsection{Center of the ER}\label{Subsection: Center of the ER}

\begin{figure}
\centering
\resizebox{\hsize}{!}{\includegraphics{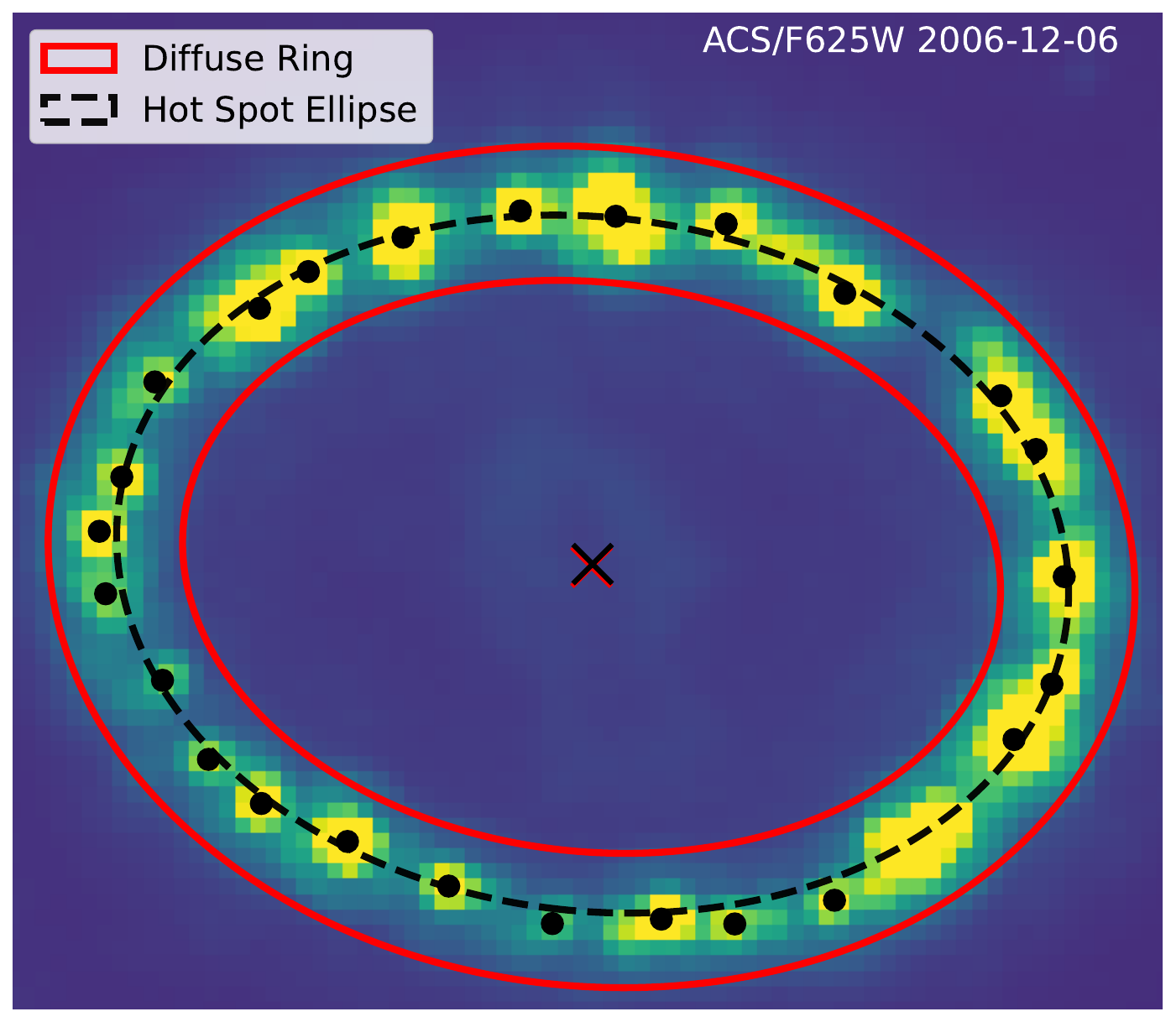}}
\caption{HST image of the ER at day 7226. The red annulus represents the fitted diffuse ring model, plotted with a width of  2$\sigma$. The black points represent the fitted hot spot centroids, while the dashed black ellipse represents the fitted ellipse around the hot spot centroids. Both methods give an estimate of the center of the ER, showcased as a black and red cross mark for the hot spot ellipse and diffuse ring, respectively.}
\label{Fig: Methods Diffuse Ring and Hot Spot Ellipse}
\end{figure}

The emission budget of the ER is H$\alpha$--dominated, with diffuse radiation being the main contributor at early epochs ($\sim2800$--$5000$~days), and higher--density hot spot emission taking over in subsequent epochs. We determined the center of the ER in the HST/WFPC2, ACS, and WFC3 images using two independent methods, with detailed mathematical descriptions provided in \cite{2024ApJ...976..164T}. In the first method, we fitted ellipses to the positions of the optical hot spots of the ER from day 6122 to day 12,978. The hot spot positions were first determined by fitting 2D elliptical Gaussians to the emission peaks. The center of the ellipse that provided the best fit to the resulting positions was then taken as the ER center. In the second method, we modeled the ER as a continuous structure using a diffuse elliptical ring with a Gaussian radial profile from day 2270 to day 12,978, which returns the center of the ER as one of its parameters. Figure~\ref{Fig: Methods Diffuse Ring and Hot Spot Ellipse} illustrates the procedure for both methods for one epoch (day 7226). Furthermore, we estimated the position uncertainties of the fitted centroids from the covariance matrix of the fits. These errors were later propagated by Monte Carlo along with the image registration uncertainties to estimate the final individual errors of the positions.

\subsection{Image registration}\label{Subsection: Image Registration}

We transformed the pixel coordinates into sky coordinates by registering the images onto the Gaia~DR3. The registration processing steps were as follows. We created a catalog of reference stars in the FOV of SN~1987A by querying Gaia DR3. We built a catalog of detected sources using the \texttt{photutils} \texttt{DAOStarFinder} tool \citep{photutils_1.9} for each of the observations. We then cross--matched the reference stars to the detected sources using a nearest--neighbor matching algorithm available in \texttt{Astropy}. We finally used the cross--matched pairs to perform the astrometric calibration based on a modified version of the \texttt{fit\_wcs\_from\_points} function from \texttt{Astropy} \citep{astropy:2013, astropy:2018, astropy:2022} that incorporates weighted least squares fitting. We opted for a general fit that includes scaling, rotation, shifts, and skewness. We estimated the image registration error of each position based on the covariance matrix of the best--fit parameters of the World Coordinate System (WCS).

Only a limited number of reference stars were detected in the $11 \times 11 $ arcsec$^{2}$ FOV of the FOC observations (see Fig.~\ref{Fig: Observations} and Table~\ref{Tab: star registration}). In the F175W exposures specifically, we found only three reference stars, which is the minimum number required to uniquely define a WCS solution. In contrast, the number of reference stars detected in the later images ranged from 10 to 30 for WFPC2, 20 to 30 for ACS, and 10 to 70 for WFC3. Most of these variations were due to differences in the FOV. Stars 2 and 3 were masked and excluded from the registration process to avoid saturation--related distortions.

We initially performed the image registration without applying proper motion corrections to the Gaia reference stars. The proper motions of the stars in our FOV are relatively small (1--5~$\rm{mas}~\rm{yr}^{-1}$), and the time intervals between the Gaia reference epoch (J2016) and our observations are less than 30 years. Under these conditions, the cross--matching procedure remains reliable provided that the search radius is appropriately adjusted. Nevertheless, to validate our results, we repeated the registration by first transforming the celestial coordinates of the reference stars from the reference epoch of Gaia~DR3 to the epoch of each observation, using the propagation functions \texttt{EPOCH\_PROP\_POS} and \texttt{EPOCH\_PROP\_ERROR} available in Gaia~DR3, which account for the proper motions published in Gaia~DR3. 

\subsection{Proper motion estimate}

To characterize the time evolution of the position of SN~1987A, we fitted its astrometric trajectory using the positions estimated at the observational epochs, derived from the registration of the images corrected for stellar proper motions. We used the \texttt{apply\_space\_motion} method available in \texttt{Astropy} to model the motion of SN~1987A and performed weighted least squares minimization with the inverse covariances as weights. We solved for the position and proper motion at the J2016 reference epoch, assuming a radial distance of 49.6~kpc \citep{2019Natur.567..200P} and a radial velocity of $287 ~\rm{km}~\rm{s}^{-1}$ \citep{2008A&A...492..481G}. Although these values may not precisely match the true distance and velocity at J2016, we verified that setting the radial velocity to zero yields an identical position for SN~1987A and induces relative changes in the proper motion of only $\sim3\%$. Given that the systemic velocity of the LMC is approximately $270 ~\rm{km}~\rm{s}^{-1}$ \citep{2024MNRAS.529.2611L}, the adopted parameters are sufficiently accurate for this analysis. This procedure allowed us to verify that the predicted position of SN~1987A at J2016 is consistent with that obtained from the registration without proper motion corrections.

\section{Results}\label{Results}

\begin{figure*}
\centering
\resizebox{\hsize}{!}{\includegraphics{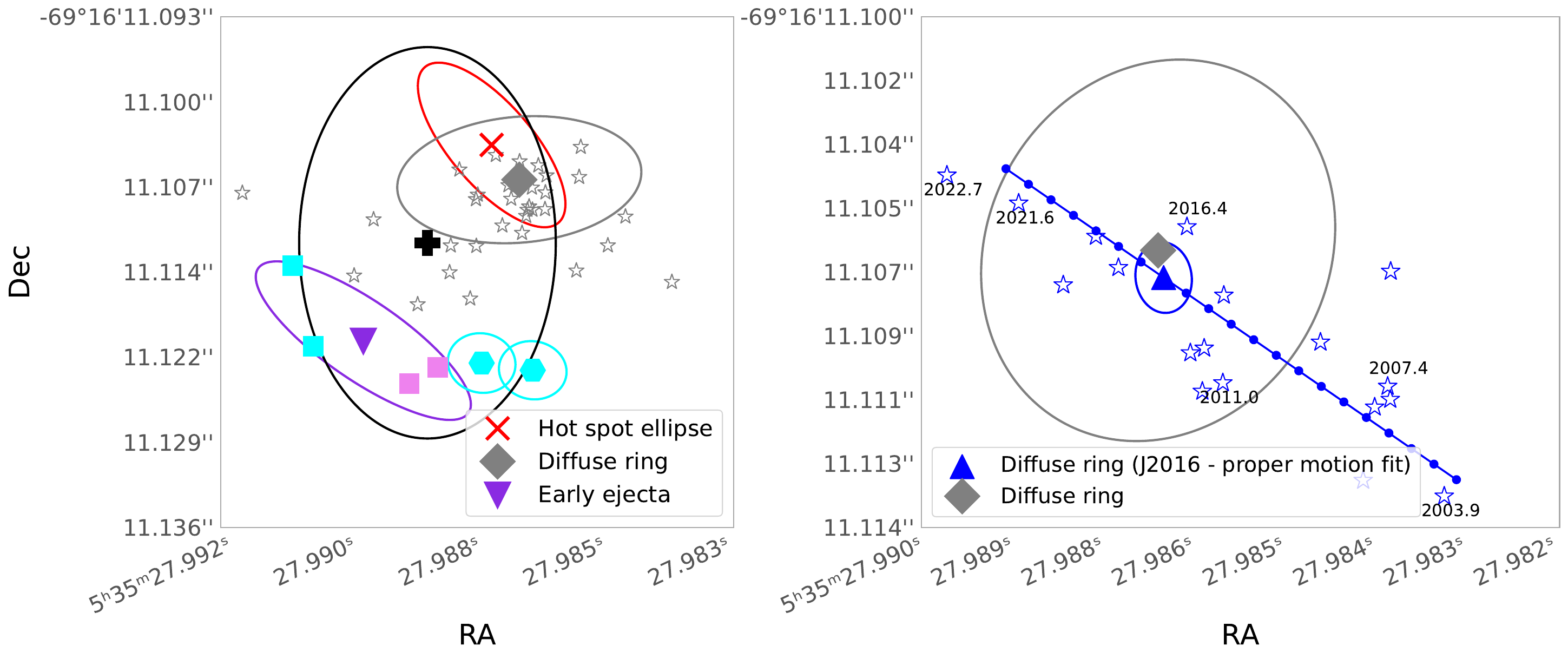}}
\caption{SN~1987A position estimates and astrometric trajectory. Left panel: Celestial coordinates of SN~1987A determined from the HST images. The gray hollow stars correspond to the diffuse ring positions after image registration onto the Gaia~DR3 frame, without correcting for the proper motion of the reference stars. The gray diamond shows the weighted average of the diffuse ring positions. The red ``$\times{}$'' marker and its ellipse are the weighted average of the hot spot ellipse positions and the 68.3\% confidence region, respectively (individual points omitted). The pink and cyan squares show the F175W (X0C80102T, X0C80103T) and F501N (X0C80106T, X0C80107T) ejecta positions, respectively. The cyan hexagons are the F501N (X0C80106T, X0C80107T) diffuse ring positions with their registration and fitting uncertainties. The upside--down purple triangle is the average of the four early ejecta positions. The black cross marks the final position, derived as the average of the mean diffuse ring and early ejecta position. The black ellipse shows the total uncertainty, corresponding to their positional discrepancy. Right panel: Time evolution of the average diffuse ring position at each epoch. The blue hollow stars show the diffuse ring center at each observational epoch, which are obtained after registering the images by accounting for the proper motion of the reference stars. For clarity, the early WFPC2 points are omitted due to their larger uncertainties and small contribution to the fit. The blue triangle marks the J2016 position from the best--fit astrometric trajectory, together with its 68.3\% confidence region. The gray diamond and its ellipse are the same as in the left panel but appear different due to the differing aspect ratios. The blue circles show the predicted positions from the best-fit trajectory of SN~1987A annually (from J2003 to J2023). The solid segments connect these points and represent the predicted proper motion from the best--fit trajectory.}
\label{Fig: proper motion}
\end{figure*}

Figures~\ref{Fig: proper motion},~\ref{Fig: result image 2016} and Table~\ref{Tab: Results} summarize the positions and correlations ($\rho^{\rm{Right~Ascension}}_{\rm{Declination}}$) of the early ejecta and the center of the ER. Unless otherwise stated, all reported error ellipses represent confidence regions at the 68.3\% level, from which the $1\sigma$ uncertainties (i.e., sample standard deviations) are derived. The final uncertainties are also based on the sample standard deviations, which should capture both the uncertainties in the fits and the image registrations.

The arithmetic mean of the four early ejecta coordinates is $\alpha = 5^{\rm{h}}~ 35^{\rm{m}}~ 27^{\rm{s}}.9899(14)$, $\delta=-69^{\circ}~ 16'~ 11''.1202(44)$ (ICRS J2016). There is an apparent systematic uncertainty of $\sim 6$ mas in the north--south direction and $\sim 12$ mas in the east--west direction between the position in the F175W and F501N filters (left panel of Fig.~\ref{Fig: proper motion}). The registration uncertainty of the individual ejecta positions in the F175W images is difficult to quantify, and we can only estimate the fitting uncertainty, which is approximately 1~mas in both coordinates. However, in the F501N images, we find that the combined fitting and registration uncertainties are less than 2~mas in both coordinates, suggesting that both uncertainties are of comparable magnitude. Nevertheless, the sample standard deviation of the four ejecta position estimates is greater than the individual uncertainties (approximately 8~mas in right ascension and 4~mas in declination).

The weighted mean of the hot spot ellipse coordinate is ${\alpha = 5^{\rm{h}}~ 35^{\rm{m}}~ 27^{\rm{s}}.9875(9)}$, ${\delta=-69^{\circ}~ 16'~ 11''.1037(46)}$ (ICRS J2016). The weighted mean of the diffuse ring coordinate is located 3~mas to the south and 3~mas to the west from that of the hot spot ellipse--an offset comparable to the uncertainties of each estimate, whose error ellipses overlap even at the 7\% confidence level. Fitting the diffuse ring in the same set of observations as the hot spot ellipse (day 6122 to day 12,978) yields a similar position offset of 2.5~mas to the south and 3~mas to the west relative to the hot spot ellipse coordinate ($\alpha = 5^{\rm{h}}~ 35^{\rm{m}}~ 27^{\rm{s}}.9870$, ${\delta=-69^{\circ}~ 16'~ 11''.1063}$). Although the earliest WFPC2 positions deviate from the bulk of the measurements (see gray stars outside the error ellipse in Fig.~\ref{Fig: proper motion}), their uncertainties are much larger and therefore they do not affect the weighted mean. The combined registration and fitting uncertainties of the individual hot spot ellipse and diffuse ring positions in both coordinates are typically 5~mas for WFC3, 5--7~mas for ACS, and 10--15~mas for WFPC2. These errors are typically 30--40\% larger for ACS and WFPC2 when proper motions are taken into account.

We adopt as our favored position of SN~1987A the average of the mean early ejecta and diffuse ring coordinates: $\alpha = 5^{\rm{h}}~ 35^{\rm{m}}~ 27^{\rm{s}}.9884(30)$, $\delta=-69^{\circ}~ 16'~ 11''.1134(136)$. We further assume that the uncertainty is represented by the discrepancy between the two independent measurements. We exclude the hot spot ellipse centroid from the final position estimate, as it lies very close to the diffuse ring center and relies on the same assumption--that the explosion occurred at the center of the ER.

The diffuse ring positions obtained from the proper motion--corrected registrations are shown as blue hollow stars in the right panel of Fig.~\ref{Fig: proper motion}. From the best--fit astrometric trajectory of SN~1987A, we derive the position of the diffuse ring at the Gaia~DR3 reference epoch (J2016) to be ${\alpha=5^{\rm{h}}~ 35^{\rm{m}}~ 27^{\rm{s}}.9869(3)}$, ${\delta=-69^{\circ}~ 16'~ 11''.1073(9)}$. The corresponding 68.3\% confidence ellipse, constructed from the covariance of the trajectory fit, overlaps with that of the hot spot ellipse position estimated without proper motion corrections. The weighted RMS of the fit residuals is 7~mas in right ascension and 5~mas in declination. Furthermore, the proper motion of the diffuse ring position at J2016, also obtained from the same trajectory fit, is ${\mu_{\rm east} (\equiv \rm{PM_{\alpha *}}) = 1.60 \pm 0.15~ \rm{mas~ yr^{-1}}}$ and ${\mu_{\rm north} (\equiv \rm{PM_{\delta}}) = 0.44 \pm 0.09~ \rm{mas~ yr^{-1}}}$, consistent with the typical proper motion field of the LMC around SN~1987A \citep{2013ApJ...764..161K, 2014ApJ...781..121V, 2016ApJ...832L..23V}. The predicted positions of SN~1987A from J2003 to J2023, based on the fitted trajectory, are shown as blue circles in Fig.~\ref{Fig: proper motion}. These positions are connected by solid segments that represent the proper motion over each one--year interval. The slope of these segments varies slightly due to the time-dependent nature of the proper motion, but this variation is only a few nanoarcseconds per year over the $\sim$30--year span, several orders of magnitude lower than the astrometric precision of our measurements.

\begin{table*}
\caption{Estimated position of SN~1987A at epoch J2016}
\label{Tab: Results}
\centering ´
\begin{tabular}{l l l l} 
\hline\hline 
&\multicolumn{3}{c}{ICRS J2016}\\ 
Method & Right ascension & Declination &$\rho^{\rm{Right~ascension}}_{\rm{Declination}}$\\
\hline 
Early ejecta 
& $5^{\rm{h}}~ 35^{\rm{m}}~ 27^{\rm{s}}.9899(14)$
& $-69^{\circ}~ 16'~ 11''.1202(44)$
& 0.82
\\
Hot spot ellipse 
& $5^{\rm{h}}~ 35^{\rm{m}}~ 27^{\rm{s}}.9875(9)$
& $-69^{\circ}~ 16'~ 11''.1037(46)$
& 0.72
\\
Diffuse ring 
& $5^{\rm{h}}~ 35^{\rm{m}}~ 27^{\rm{s}}.9869(15)$
& $-69^{\circ}~ 16'~ 11''.1066(35)$
& -0.12
\\
\hline
Final position\tablefootmark{a}
& $5^{\rm{h}}~ 35^{\rm{m}}~ 27^{\rm{s}}.9884(30)$
& $-69^{\circ}~ 16'~ 11''.1134(136)$
& 0.00
\\
\hline
\end{tabular}
\tablefoot{\\
\tablefoottext{a}{Average of diffuse ring and early ejecta position.}
}
\end{table*}

In contrast, the trajectory of the hot spot ellipse centroid is less systematic: ${\mu_{\rm east} = 1.26 \pm 0.15~\rm{mas~yr^{-1}}}$ and ${\mu_{\rm north} = -0.02 \pm 0.10~\rm{mas~yr^{-1}}}$, with the latter consistent with zero (relative uncertainty of 500\%). However, the predicted J2016 position from this trajectory agrees with the no--proper--motion estimate to within 2~mas in both coordinates. The less systematic nature is most likely caused by flux evolution within the ER, which introduces centroid shifts unrelated to true proper motion. The hot spot ellipse method is more sensitive to such fluctuations because it fits fewer points than the diffuse ring method. To mitigate these variations, we explored a workflow in which all epochs are first resampled onto a common astrometric frame and hot spot positions are averaged across epochs before registering to Gaia~DR3. In this configuration, the trajectory is more coherent, yielding ${\mu_{\rm east} = 2.02 \pm 0.20~\rm{mas~yr^{-1}}}$ and ${\mu_{\rm north} = 0.85 \pm 0.12~\rm{mas~yr^{-1}}}$.

\section{Discussion}\label{Discussion}

\begin{figure*}
\centering
\resizebox{\hsize}{!}{\includegraphics{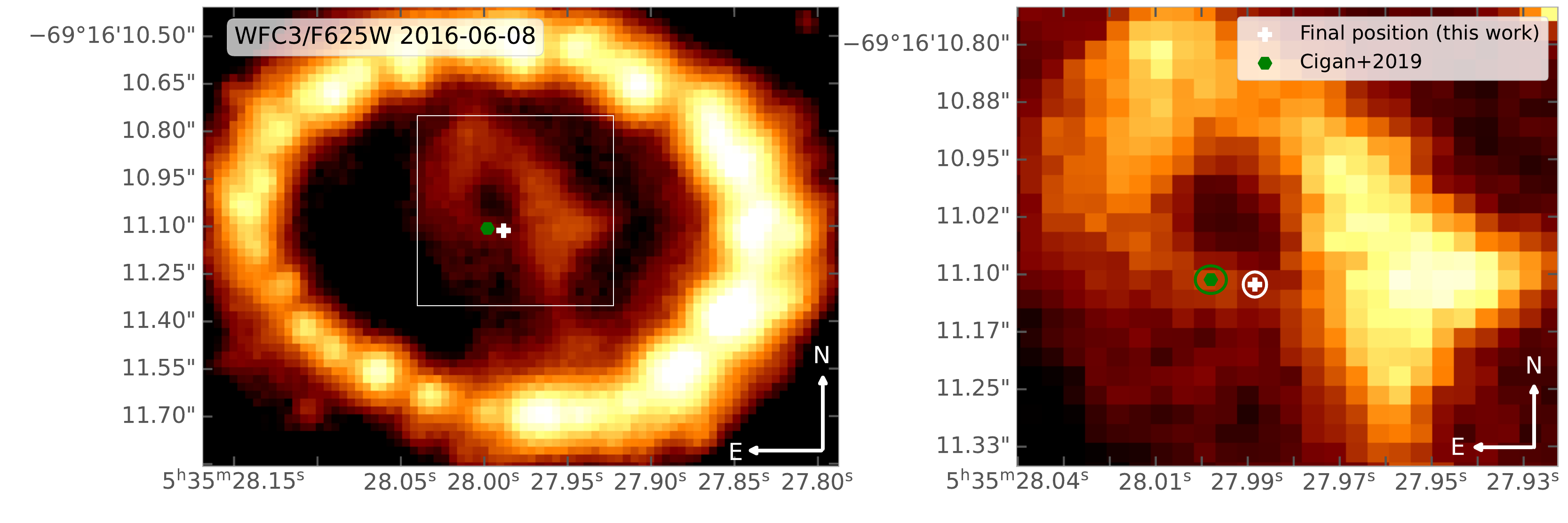}}
\caption{HST/WFC3/F625W image of the ER and ejecta at day 10,698. The image is registered onto the Gaia DR3 frame (J2016). Left panel: Full view of the ER showing the favored final position of SN~1987A and the ALMA submillimeter position from \citet{2019ApJ...886...51C}. The FOV is $2\farcs025 \times 1\farcs450$. The white rectangle outlines the region shown in the right panel. Right panel: Zoom-in on the ejecta with the final position and the ALMA estimate overlaid. The FOV is  $0\farcs625 \times 0\farcs600$.}
\label{Fig: result image 2016}
\end{figure*}

\begin{figure}
\centering
\resizebox{\hsize}{!}{\includegraphics{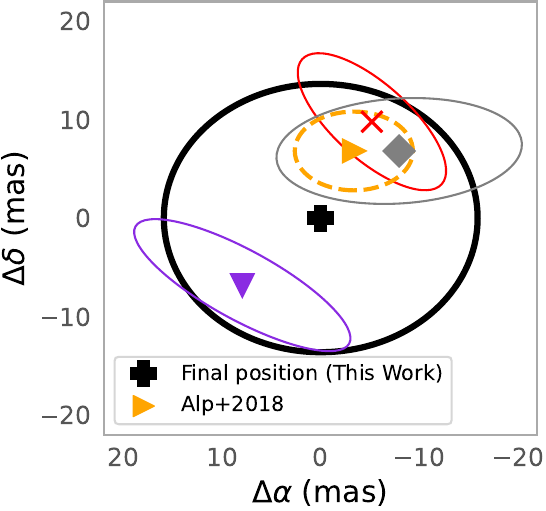}}
\caption{Individual positions relative to final explosion center. The black cross shows the favored final position $\alpha=5^{\rm{h}}~ 35^{\rm{m}}~ 27^{\rm{s}}.9884(30)$, $\delta=-69^{\circ}~ 16'~ 11''.1134(136)$ (ICRS J2016). The black circle indicates the total uncertainty, defined as the offset between the mean early ejecta and mean diffuse ring positions. The upside-down purple triangle marks the average early ejecta position; the red ```$\times{}$'' marker shows the average hot spot ellipse position; the gray diamond indicates the average diffuse ring position. Their corresponding ellipses represent 68.3\% confidence regions. The right orange triangle shows the position from \citet{2018ApJ...864..174A}, corrected from J2015 ($5^{\rm{h}}~ 35^{\rm{m}}~ 27^{\rm{s}}.9875(11)$, $-69^{\circ}~ 16'~ 11''.1070(40)$) to J2016 ($5^{\rm{h}}~ 35^{\rm{m}}~ 27^{\rm{s}}.9878$, $-69^{\circ}~ 16'~ 11''.1066$) using the estimated proper motion of SN~1987A.}
\label{Fig: final position}
\end{figure}

The estimated centers of the early ejecta and the ER differ such that their error ellipses overlap at the 96\% confidence level (error ellipses that correspond to a chi--squared quantile of $q\sim 6.4$ with 2 degrees of freedom). There are several different uncertainties that could contribute to this discrepancy, including differences between the instruments and filters used for the two measurements. Such differences are observed even between the \textit{R}--band filters of WFPC2, ACS, and WFC3 exposures, as is illustrated by both panels in Fig.~\ref{Fig: proper motion}. Finally, we note that the registration of the FOC images is based on a limited number of reference stars compared to the later HST observations, which can introduce further systematic uncertainties.

Another possibility is that the explosion itself might have been offset from the center of the ER. We further explore this scenario by fitting the diffuse ring model to the barely visible ER in the early F501N images. The average of the two positions, shown as cyan hexagons in the left panel of Fig.~\ref{Fig: proper motion}, is $\alpha = 5^{\rm{h}}~ 35^{\rm{m}}~ 27^{\rm{s}}.9872$, $\delta=-69^{\circ}~ 16'~ 11''.1224$. The early ejecta positions of the same registered exposures lie 21~mas east and 5~mas north of this estimate, an offset that is relatively large compared to the statistical uncertainties of the ER positions shown in Fig.~\ref{Fig: proper motion}. This indicates that there may indeed be a small, real offset between the center of the explosion and the center of the ER. However, we note that the results may also be biased by the fact that the F501N filter is narrow and does not capture all the emission from the ejecta, which emits broad emission lines.
The simplest explanation for a possible real offset between the two positions is that the mass loss that created the ER was not entirely symmetric around the progenitor. This is supported by the variations in surface brightness across the ER, with the southeast side exhibiting the faintest emission, especially at late epochs \citep{2024ApJ...976..164T}. Such azimuthal flux variations can bias the centroids toward the brighter side, which is consistent with the direction of the offset between the ejecta and ER centroids. The most important consequence of the offset is the impact on the inferred kick velocity of the neutron star, discussed further below.

Our best position can be compared to previous estimates, see Figs.~\ref{Fig: result image 2016} and \ref{Fig: final position}. \citet{2019ApJ...886...51C} derived a ring center position of $\alpha = 5^{\rm{h}}~ 35^{\rm{m}}~ 27^{\rm{s}}.998$, $\delta=-69^{\circ}~ 16'~ 11''.107$ (ICRS\footnote{No epoch was provided, but we take it to be J2015.5, which corresponds to the epoch of the ALMA observation.}). Their estimate was obtained by fitting the ring emission in an ALMA observation obtained in 2015 at 315~GHz, using three independent fitting methods. The final position was taken as the average of these methods, and the reported uncertainties were combined from both the fitting procedure and the astrometric calibration (18~mas in both coordinates). We note that the positional offset relative to our estimate (52~mas in right ascension and 6~mas in declination) is not surprising because of wavelength--dependent differences in the origin of the emission. In the ALMA image, the emission from the ER is dominated by synchrotron radiation from the reverse shock. The reverse shock extends into low--density regions at high latitudes above and below the plane of the ER (see \citealp{2023ApJ...949L..27L}), implying that the observed morphology is influenced by projection of emission from a large region. In contrast, the optical emission observed in the HST images primarily originates from higher--density gas confined to the ER itself and should therefore provide a more reliable estimate of the center. 

Figure~\ref{Fig: final position} also shows the position inferred by \citet{2018ApJ...864..174A} for direct comparison with our result. They report a position of $\alpha=5^{\rm{h}}~ 35^{\rm{m}}~ 27^{\rm{s}}.9875(11)$, $\delta=-69^{\circ}~ 16'~ 11''.1070(40)$ (ICRS J2015), determined from fitting ellipses to the locations of the hot spots. The small offset relative to our hot spot center can be explained by the combination of several small differences in the methods and data used for the analysis. The estimate in \citet{2018ApJ...864..174A} was based on HST/WFPC2, ACS, and WFC3 \textit{B}-- and \textit{R}--band observations obtained between 2003 and 2016, and was tied to the Gaia DR1 reference frame. In addition, \citet{2018ApJ...864..174A} used 1D Gaussian profiles to estimate the positions of the hot spots around the ER while our method involved fitting 2D Gaussian functions instead.

The uncertainty in the position of the center directly affects the estimate of the kick velocity of the neutron star, which relies on a precise knowledge of the center of explosion. If the small offset between the ejecta and ER positions is due to asymmetries in the mass loss affecting the latter, it follows that the early ejecta centroid should provide a better estimate of the center of explosion. At the same time, our estimated ejecta centroid has a much larger statistical uncertainty and may also be biased by the use of the narrow F501N filter, which does not capture all the ejecta emission. Asymmetries in the explosion may also produce an offset between the peak of the ejecta emission and the explosion center, with the ejecta centroid moving in the opposite direction to the neutron star due to momentum conservation. However, this is expected to have a negligible impact on our results given the large difference between the masses of the ejecta and neutron star, which implies a slow movement of the centroid, in combination with the observations being obtained at early times, only three years after the explosion. Given all these uncertainties, we consider the average of the ejecta and ER positions as the best estimate of the explosion center, and take the offset of 21~mas as a systematic uncertainty. 

The compact object in SN~1987A was recently identified in JWST data based on the detection of emission lines from  [Ar~II], [Ar~VI], [S III-IV] and [Ca~V], originating from a spatially unresolved region close to the center of the remnant \citep{2024Sci...383..898F,2025Larsson}. This emission originates from ejecta that have been ionized by the compact object, as demonstrated by the photoionization models in \cite{2024Sci...383..898F} and \cite{2025Larsson}. However, it is currently unclear whether this object is just a thermally emitting neutron star or whether it also harbors a pulsar wind nebula (PWN), which introduces an additional uncertainty in the distance between the neutron star itself and the observed emission region. If this distance is small, the observed offset between the emission region and the center of the explosion ($d$) provides a direct measure of the kick velocity in the sky plane, $v_{\rm kick, sky} = d/t$, where $t$ is the time since explosion. The assumption of a small distance between the neutron star and emission region is likely good in the scenario with a thermally emitting neutron star, while it is more uncertain in the PWN scenario (see \citealt{2025Larsson} for further discussion).

The most precise measure of the JWST source is provided by the [Ar~VI]~4.5292~$\mu$m line, which is located $30\pm 10~ \rm{mas}$ south and $63 \pm 10~ \rm{mas}$ east of the ER center reported by \cite{2018ApJ...864..174A} \citep{2025Larsson}.  This translates to a kick velocity in the sky plane of $443\pm 64\ \rm{km\ s^{-1}}$. As the line is also blueshifted by $252.3\pm 1.8\ \rm{km\ s^{-1}}$, the total inferred 3D kick velocity is $510\pm 55\ \rm{km\ s^{-1}}$, where the uncertainty is statistical only. Taking the new favored position obtained in this work as the explosion center has a very minor impact on this result, giving a kick in the sky plane of $399\pm148\ \rm{km\ s^{-1}}$, and a total 3D kick of $472\pm126\ \rm{km\ s^{-1}}$, where we have now also included the systematic uncertainty in the explosion center in the confidence interval. By contrast, taking the early ejecta position as the center reduces the kick in the sky plane to $337\pm 148\ \rm{km\ s^{-1}}$, giving a total 3D kick of $420\pm119\ \rm{km\ s^{-1}}$. Given the relative location of these three centers (Fig.~\ref{Fig: final position}), the inferred kick in the sky plane is always directed to the southeast with a position angle in the range 109--116$^{\circ}$, defined counterclockwise from the north.

The magnitude of the kick in SN~1987A is comparable to the typical 3D velocity of $\sim 400\ \rm{km\ s^{-1}}$ estimated from studies of radio pulsars \citep{2005Hobbs,2006FaucherGiguere}, though a detailed comparison with the pulsar population is not possible due to the relatively large uncertainty on the kick. Reducing this uncertainty would also be important for constraining models for the explosion mechanism. At present, the biggest uncertainty affecting the inferred kick velocity is the question of whether there is a PWN. This question is expected to be resolved with future observations as the optical depth to dust scattering and absorption decreases quadratically with time as the ejecta expand. The uncertainty in the center of the explosion will then become the dominant uncertainty in the kick velocity. Given that the determination of the center of the explosion is partly based on early observations, it is unlikely that this systematic uncertainty will be reduced in future studies. However, the related uncertainty on the kick velocity in the sky plane will decrease with time as the neutron star travels further from the center, following the relation $123~ (t/40~ \rm{yr})^{-1}~ \rm{km~s}^{-1}$, where $t$ is the time since the explosion. Future observations are therefore key for determining both the nature of the compact object and its kick velocity. 

\begin{acknowledgements}
      This work was supported by the Knut \& Alice Wallenberg foundation.
\end{acknowledgements}

\bibliographystyle{aa}
\bibliography{ref}

\end{document}